\documentclass[prl,showpacs,twocolumn]{revtex4}

\newcommand{\ket}[1]{|{#1}\rangle}

\newcommand{\be}{\begin{equation}}
\newcommand{\ee}{\end{equation}}

\newcommand{\llt}{\ll}

\usepackage{amsfonts}
\usepackage{amsmath}
\usepackage{rotating}
\usepackage{graphicx}
\usepackage{amssymb}
\usepackage{amsmath}
\usepackage{amssymb}
\usepackage{graphicx}
\usepackage{lscape}
\usepackage{color}
\usepackage{epstopdf}

\begin{document}
\title{Few-Body Route to One-Dimensional Quantum Liquids}

\author{Manuel Valiente}
\affiliation{SUPA, Institute of Photonics and Quantum Sciences, Heriot-Watt University, Edinburgh EH14 4AS, United Kingdom}
\author{Patrik \"Ohberg}
\affiliation{SUPA, Institute of Photonics and Quantum Sciences, Heriot-Watt University, Edinburgh EH14 4AS, United Kingdom}
\begin{abstract}
Gapless many-body quantum systems in one spatial dimension are universally described by the Luttinger liquid effective theory at low energies. Essentially, only two parameters enter the effective low-energy description, namely the speed of sound and the Luttinger parameter. These are highly system dependent and their calculation requires accurate non-perturbative solutions of the many-body problem. Here, we present a simple method that only uses collisional information to extract the low-energy properties of these systems. Our results are in remarkable agreement with available results for integrable models and from large scale Monte Carlo simulations of one-dimensional helium and hydrogen isotopes. Moreover, we estimate theoretically the critical point for spinodal decomposition in one-dimensional helium-4, and show that the exponent governing the divergence of the Luttinger parameter near the critical point is exactly 1/2, in excellent agreement with Monte Carlo simulations.
\end{abstract}
\maketitle


\paragraph{Introduction}. Interacting quantum systems in one spatial dimension, long ago considered toy models far away from the three-dimensional reality, now hold the status of physically relevant theories. Advances in the transversal confinement of trapped ultracold atomic gases \cite{BlochReview,Haller}, the realisation of carbon nanotubes by rolling up sheets of graphene \cite{Saito,Charlier}, or helium isotopes adsorbed in nanopores \cite{Yager,DelMaestro}, make it possible to investigate many-body quantum physics in wire geometries with unprecedented level of control. Most one-dimensional systems, whether weakly or strongly interacting, are universally described by the Luttinger liquid effective field theory at low energies \cite{HaldaneJPC}, and by its recently developed non-linear counterpart at higher energies \cite{Imambekov}. Essentially, many correlation functions, and the excitation spectrum, have universal behaviours, and the non-universal parameters -- the Luttinger parameter and speed of sound  -- are the only system-dependent quantities of interest.  To extract these, however, one needs to either invoke perturbation theory, only valid for weak interactions, or to solve the many-body problem numerically "exact" using Monte Carlo \cite{DelMaestro,Bertaina,AstrakharchikHe,AstrakharchikH} for continuous or DMRG methods \cite{Schollwock} for lattice models, or quasi-analytically for integrable models via the Bethe ansatz \cite{Cazalilla,Imambekov}. In this Letter, we develop a simple, yet highly non-perturbative method, that uses only two-body scattering information to extract the speed of sound and Luttinger parameter of strongly interacting many-body quantum systems in one dimension. To show the reliability of our theory, we study all the stable isotopes of helium and spin-polarised hydrogen, and tritium, when tightly confined to one dimension, using realistic molecular potentials, which are strongly-interacting and intractable with perturbative methods, and compare our results to the Monte Carlo data of references \cite{Bertaina,AstrakharchikHe,AstrakharchikH}. Using similar methods, we also study the liquid phase of $^4\mathrm{He}$, which is not a Luttinger liquid.

The excitations of gapless one-dimensional many-body systems are characterised, at low energy, by the speed of sound $v$, corresponding to an excitation spectrum $\hbar \omega(q)=\hbar v q + \mathcal{O}(q^2)$, with $q$ the momentum of the excitation. For bosons, for instance, in the weak and strong coupling limits, corresponding to the quasi-condensate and fermionised (or Tonks-Girardeau) regimes, respectively, Bogoliubov and many-body (fermionic) perturbation theory can be used to extract approximations to the speed of sound \cite{Cazalilla,LiebLiniger,Minguzzi} in these limits. A weak-coupling approximation to the speed of sound can also be obtained by fitting Tomonaga-Luttinger model's coupling constant to reproduce the correct reaction matrix of the original model in the fermionised situation \cite{ValientePhillipsZinnerOhberg}. Unfortunately, no simple and reliable non-perturbative approximation is currently available. 

One-dimensional spin-polarised Fermi gases (or strongly-coupled spinless Bose gases) owe much of their special phenomenology to the fact that the Fermi surface is composed of only two (Fermi) points $\pm k_F$. It is reasonable to assume that one can go beyond first order (Born) approximation in the speed of sound by using only two particles that, in the case of no interactions, have momenta $\pm k_F$, based on the fact that $N$-th order perturbation theory only needs processes involving $N$ "active" particles close to the Fermi points. The complexity of many-body perturbation theory, as opposed to few-body physics, is much increased by the presence of the Fermi sea. Here, we strive to obtain a non-perturbative method that uses two-body physics only, without the extra complications of the Fermi sea.

\paragraph{Two fermions on a ring}. Consider two spin-polarised fermions whose dynamics is governed by the Hamiltonian
\begin{equation}
H=\frac{p_1^2}{2m} +\frac{p_2^2}{2m} + V(x_1-x_2).\label{Ham}
\end{equation}
If the interaction $V(x)$, with $x\equiv x_1-x_2$, falls off faster than $1/|x|$ at long distances, then the stationary scattering states $\psi_k(x)$, after separation of centre of mass and relative coordinates, of Hamiltonian (\ref{Ham}), behave asymptotically as
\begin{equation}
\psi_k(x)\to \mathrm{sgn}(x) \sin(k|x|+\theta(k)),\hspace{0.1cm} x\to \pm \infty.
\end{equation}
Above, we assume $k>0$, and $\mathrm{sgn}(x)$ is the signum function. We place the two fermions on a ring of length $L$ (periodic boundary conditions). It is easy to see that, for total momentum $K=0$, the energies of the two-body states are given by $\hbar^2k^2/m$, where $k$ must satisfy the following equation
\begin{equation}
k=\frac{2\pi n}{L} - \frac{2}{L}\theta(k),\label{keq}
\end{equation}
with $n\in \mathbb{Z}^+$. From now on, we particularise to $n=1$ in Eq.~(\ref{keq}), and identify $k_F=2\pi/L$. 

\paragraph{Universality hypothesis}. In a many-body system, the thermodynamic limit is attained by taking the length of the ring to infinity while keeping the density constant. This, in a few-particle system, is not possible. To remedy this inherent deficiency, we shall invoke a universality hypothesis, stating that two thermodynamically large one-dimensional, gapless quantum systems at zero temperature but at different densities have the same dimensionless low-energy properties if all their dimensionless coupling constants at energies near the Fermi points are identical. Clearly, this hypothesis is very reasonable and is verified in well-known integrable models, such as the Lieb-Liniger model \cite{LiebLiniger}, or its fermionic dual \cite{CheonShigehara,ValienteZinner,Cui}, whose ratio of speed of sound to Fermi velocity, and its Luttinger parameter only depend on the coupling constant $\gamma$ (see supplemental material), regardless of the density of the system. For generic, non-integrable models, with two-body forces only, the universality hypothesis implies that the two-body $T$-matrices, on- and off-shell, of the two target models must be the same. The theory we present below is a pure two-particle theory, and therefore we can only invoke universality on the on-shell (i.e. phase shifts) scattering. Note that universality in this case implies that the speed of sound, when $\theta(k_F)=0$, must be given by the Fermi velocity $v=v_F$. This last condition, which is not exact, is a very a good approximation nevertheless as we will see below (see also \cite{Bertaina,AstrakharchikHe,AstrakharchikH}), and coincides with the result obtained by weak-coupling theory from the Tomonaga-Luttinger model \cite{Giamarchi,Imambekov,ValientePhillipsZinnerOhberg}. 

\paragraph{Calculation of the speed of sound}. In order to excite the system, we need to add more incident relative momentum in Eq.~(\ref{keq}). The lowest excitation is obtained by setting $n=2$, corresponding to $2k_F$, which is too large to extract low-momentum expansions of the excitation energies. Instead, we can shorten the ring, $L\to L-\delta$, with $0<\delta\llt L$. Identifying, to leading order, the incident momentum in Eq.~(\ref{keq}) with $k_F+q/2$, we obtain $\delta=L^2q/4\pi$. Since in this way not only the incident momentum ($\equiv k_F$), but also the density is increased, we need to ensure that the coupling constants of the theory, such as the gas parameter $\rho a$ for hard rods (HR) (see supplemental material), or the Lieb-Liniger (LL) parameter $\gamma$ remain fixed, in order to guarantee that the universality hypothesis stated in the previous paragraph holds at the two-body level. We find that the speed of sound is given by
\begin{equation}
\hbar v =\frac{\hbar^2L^2}{4\pi m}\frac{\mathrm{d}k_{\delta}^2}{\mathrm{d}\delta}\left.\right|_{\delta=0},\label{speedofsound}
\end{equation}
while leaving the dimensionless coupling constants of the particular problem fixed, and where $k_{\delta}$ is the solution to Eq.~(\ref{keq}) with $L$ substituted by $L-\delta$. 

\begin{figure}[h]
\includegraphics[width=0.5\textwidth]{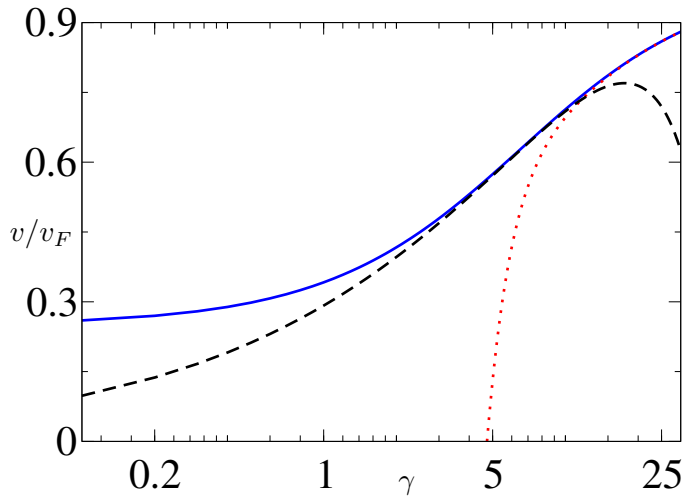}
\caption{Speed of sound vs. Lieb-Liniger constant $\gamma$. We plot  the theoretical results from Eqs.~(\ref{speedofsound}) and (\ref{keq}) (blue solid line), 4-th order perturbation theory in $1/\gamma$ (red dotted line), and Bogoliubov theory (black dashed line).}
\label{fig:BogoPerto}
\end{figure}

To show how good an approximation one can achieve with Eq.~(\ref{speedofsound}), we calculate the speed of sound for the HR (with diameter $a$) \footnote{Or the extended HR model \cite{ValienteEPL} with scattering length $a$.} and the LL model (to third order), obtaining 
\begin{align}
\frac{v_{\mathrm{HR}}}{v_F}&=(1-\rho a)^{-2}\label{vHR}\\
\frac{v_{\mathrm{LL}}}{v_F}&=1-\frac{4}{\gamma}+\frac{12}{\gamma^2}+\frac{16}{3\gamma^3}(\pi^2-6)+\mathcal{O}(\gamma^{-4}).
\end{align}
Above, and as one could expect for the very simple HR model, we see that its speed of sound is exact \cite{Bertaina,Imambekov}, while for the Lieb-Liniger model it is correct to third order in perturbation theory \cite{Minguzzi}. In Fig.~\ref{fig:BogoPerto} we plot the numerically calculated speed of sound for the LL model and compare it with 4th-order perturbation theory \cite{Minguzzi} in $1/\gamma$ and the Bogoliubov approximation \cite{Cazalilla,LiebLiniger}. There, we observe that the results are highly non-perturbative and are valid beyond the perturbative regime and down to $\gamma\approx 5$, and that our results interpolate between the Bogoliubov and the Tonks-Girardeau regimes.

\begin{figure}[t]
\begin{center}
\includegraphics[width=0.5\textwidth]{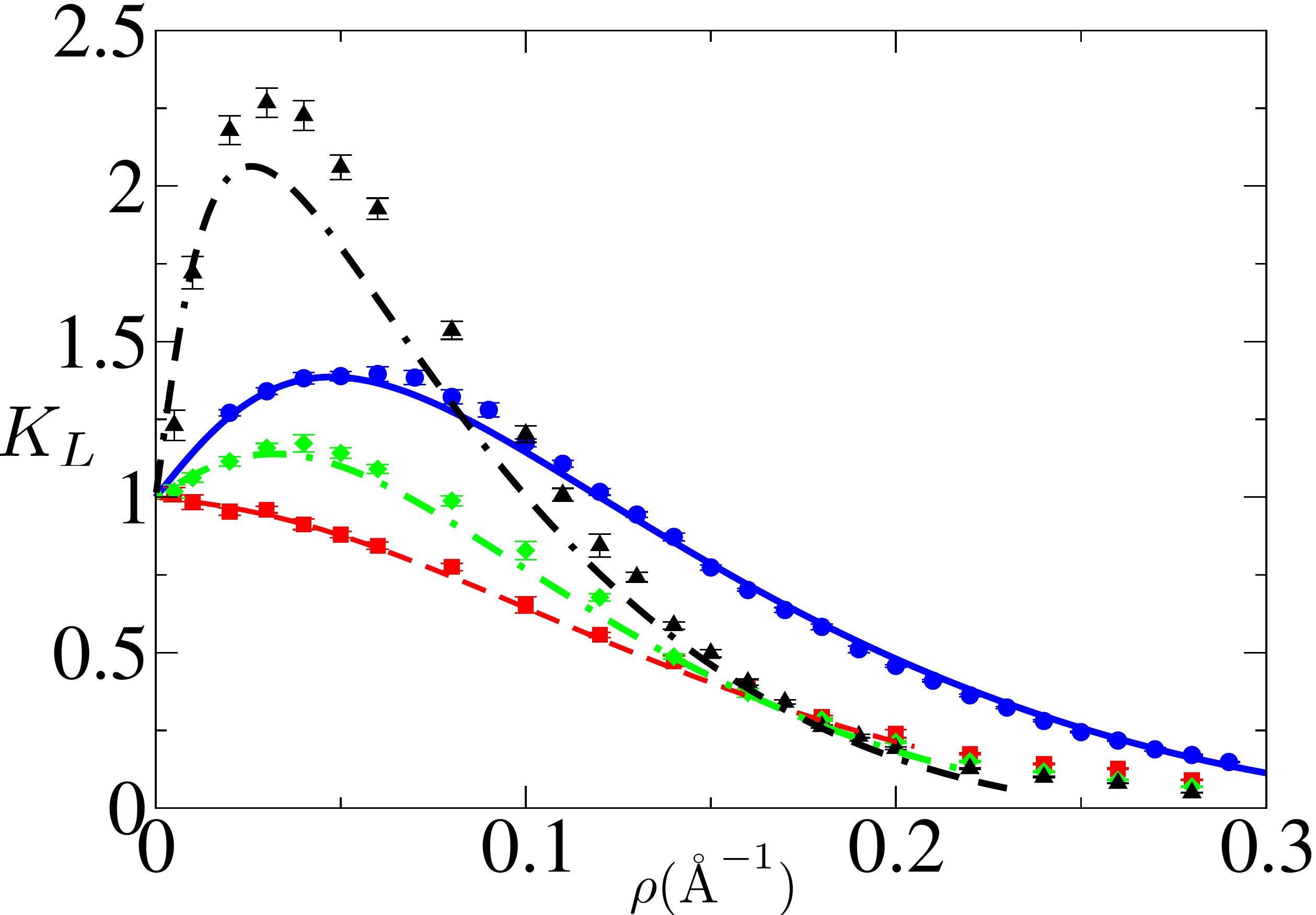}
\end{center}
\caption{Luttinger parameter vs. density for $^{3}\mathrm{He}$ (blue solid line), $^{3}\mathrm{H}$ (black double dashed-dotted line), $^{2}\mathrm{H}$ (green dashed-dotted line) and $^{1}\mathrm{H}$ (red dashed line). Blue dots ($^{3}\mathrm{He}$), black triangles ($^{3}\mathrm{H}$), green diamonds ($^{2}\mathrm{H}$) and red squares ($^{1}\mathrm{H}$) are Monte Carlo data from references \cite{AstrakharchikHe} and \cite{AstrakharchikH}.}
\label{fig:He3H}
\end{figure}

\begin{figure}[t]
\includegraphics[width=0.5\textwidth]{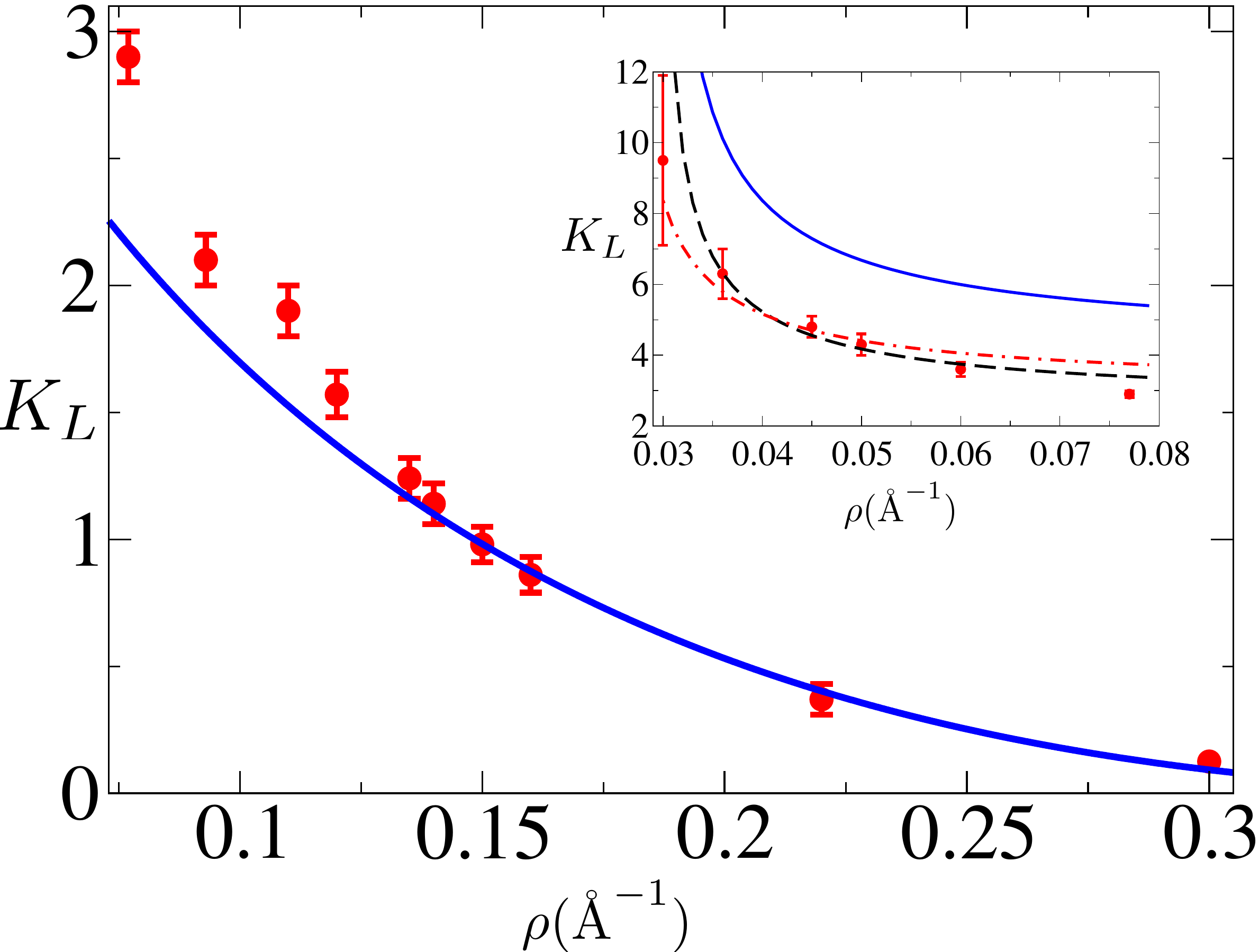}
\caption{Luttinger parameter vs. density for $^4\mathrm{He}$ (solid blue line). Red dots are Monte Carlo data of reference \cite{Bertaina}. Inset: Luttinger parameter from Eq.~(\ref{KL}) (blue solid line), renormalised Luttinger parameter (see text) with $\rho_*=0.0296\mathrm{\AA}^{-1}$ (black dashed line), and with $\rho_*=0.026\mathrm{\AA}^{-1}$ (red dashed-dotted line). Red dots are Monte Carlo data of reference \cite{Bertaina}.}
\label{fig:He4}
\end{figure}

\paragraph{Light atomic gases}. We now move on to discuss more realistic, non-integrable systems in one dimension. We study $^4\mathrm{He}$, $^3\mathrm{He}$, and spin-polarised $^3\mathrm{H}$, $^2\mathrm{H}$ and $^1\mathrm{H}$. For all the above atoms there are extensive Monte Carlo data \cite{Bertaina,AstrakharchikHe,AstrakharchikH} for the Luttinger parameter $K_L=v_F/v$ \cite{Giamarchi}. We calculate the phase shifts $\theta(k)$ using the Aziz HFDHF2 potential for $^4\mathrm{He}$ \cite{AzizHF}, the Aziz II potential for $^3\mathrm{He}$ \cite{AzizII,BoronatAziz}, and a cubic-spline interpolation of the JDW potential \cite{JDW}, readjusted to match the $1/|x|^6$ tails \cite{JDW2} as in reference \cite{AstrakharchikH}, for hydrogen. Since all these interactions have short-distance hard cores, the distinction between fermion and boson is void and we will treat all these atoms as fermions. The phase shifts (see supplemental material) can be locally approximated by linear functions as
\begin{equation}
\theta(k)\approx \theta_0-a(k-k_F),\label{linearphase}
\end{equation}
where $\theta_0$ is the phase shift at $k=k_F$. We have verified that adding higher order terms in Eq.~(\ref{linearphase}) does not change the results below in a significant way. The manipulations in Eq.~(\ref{speedofsound}) are done by keeping the dimensionless coupling constants $\theta_0$ and $k_Fa$ fixed. The results are shown in Figs.~\ref{fig:He3H} and \ref{fig:He4}, and are in excellent agreement with the Monte Carlo results of references \cite{Bertaina,AstrakharchikHe,AstrakharchikH}. As noted above, the universality hypothesis at the two-body level implies $K_L=1$ whenever $\theta(k_F)=0$ which, as seen in Figs.~\ref{fig:He3H} and \ref{fig:He4}, is nearly true in most cases. This constraint may be lifted by non-integrability effects beyond two-body physics.

\begin{figure}[t]
\begin{center}
\includegraphics[width=0.5\textwidth]{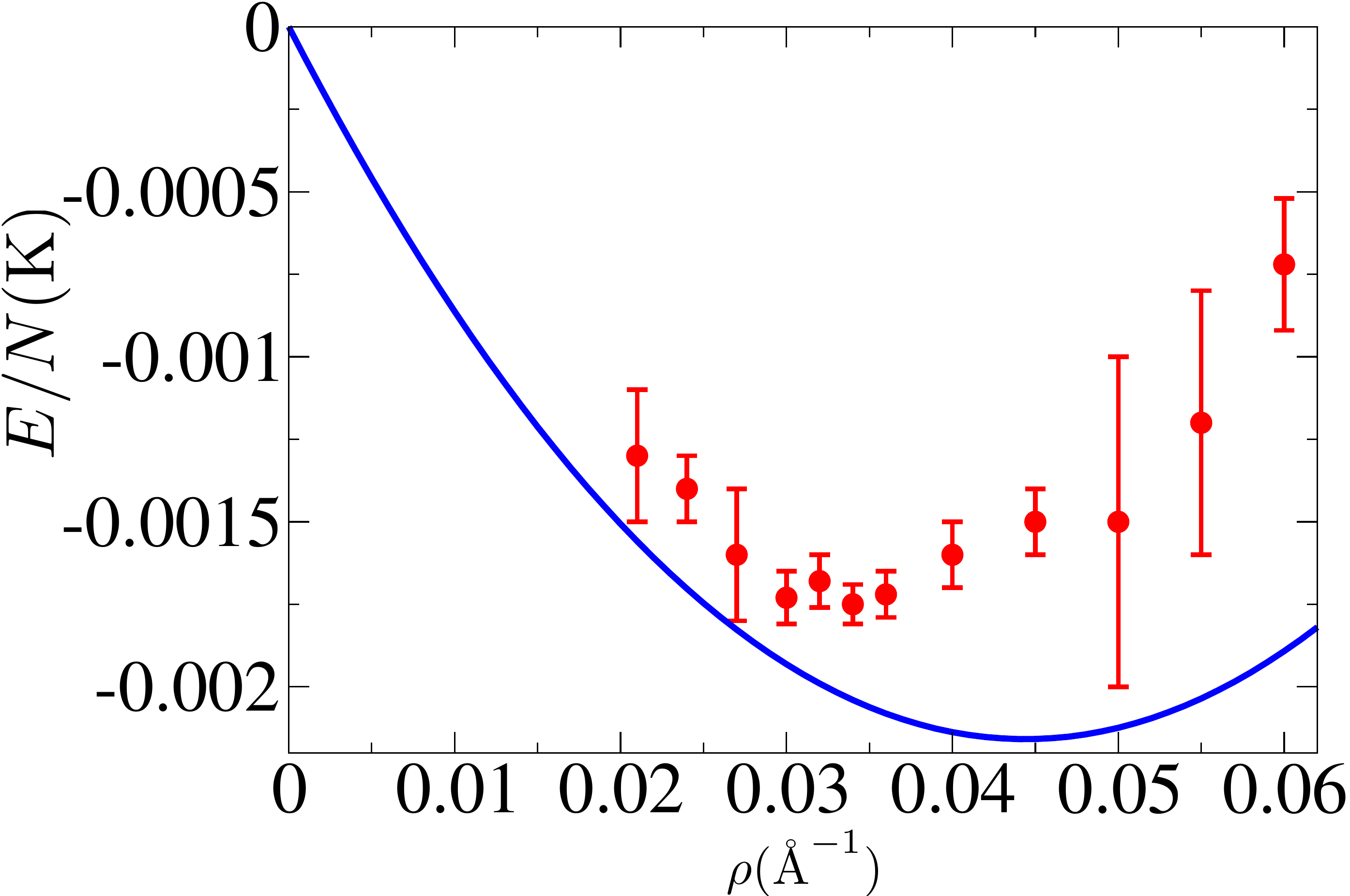}
\end{center}
\caption{Low-density equation of state from Eq.~(\ref{EoS}) (blue solid line). Red dots are Monte Carlo data of reference \cite{Bertaina}. }
\label{fig:EoS}
\end{figure}

\paragraph{Liquid $^4\mathrm{He}$}. Of all the helium and hydrogen isotopes, the most difficult to describe is $^4\mathrm{He}$, as is seen in Fig.~\ref{fig:He4}. This is due to the existence of a very weakly bound s-wave state in three dimensions, with a binding energy $E_B= 1.1 + 0.3 /- 0.2 \mathrm{mK}$ \cite{Grisenti} \footnote{Note that the binding energy of $^4\mathrm{He}_2$ is seven orders of magnitude smaller than for $\mathrm{H}_2$, for instance}. In strict one dimension, the binding energy is identical to the three-dimensional case due to the short-distance hard core of the He-He interaction. For the Aziz HFDHF2 potential, the scattering length and effective range are $a=124.65\mathrm{\AA}$ and $r=7.39\mathrm{\AA}$, respectively, yielding $E_B \approx 0.83\textrm{mK}$. The liquid phase of $^4\mathrm{He}$ \cite{Bertaina} exists at low densities $\rho$ below a critical point $\rho_*$, which was  calculated as $\rho_*=0.026\pm 0.002 \mathrm{\AA}^{-1}$ in reference \cite{Bertaina}. To estimate the critical point theoretically, we use the well-established energy-dependent scattering length $a(k)$ \cite{Blume}, given by $-1/a(k)=-1/a+rk^2/2$, where $k=\sqrt{mE/\hbar^2}$, and $E$ the relative energy, giving an effective interaction strength $g(k)=-2m/\hbar^2a(k)$, with $V\sim g(k)\delta(x)$. We place two particles in a ring of length $L$ and find the critical $k$, $k_*$, by setting $g(k_*)=0$, which yields $k_*= 0.0466\mathrm{\AA}^{-1}$. This is in very good agreement with the resonance found numerically, located at $k_*=0.0465\mathrm{\AA}^{-1}$. Identifying $\rho=2/L$, we obtain $\rho_*=2k_*/\pi=0.0296\mathrm{\AA}^{-1}$, in good agreement with reference \cite{Bertaina}. At low densities, $\rho\ll \rho_*$, we expect the effective range to contribute minimally, and we can use the attractive mean-field (Gross-Pitaevskii) theory with Lieb-Liniger parameter $\gamma=-2/\rho a$. At higher densities, the effective range will play a role, but effective-range mean-field theory with a homogeneous ground state shows no effect of $r$ \cite{Salasnich}. We therefore use a Hammer-Furnstahl field redefinition \cite{HammerFurnstahl,HammerRMP} to trade the effective range for a three-body force of strength $\lambda_3$, and avoid its microscopic calculation \cite{Kohler,Tomio} by using $\partial_{\rho} \mu=0$ at $\rho=\rho_*$, where $\mu=g\rho+\lambda_3\rho^2$ is the chemical potential. This yields $\lambda_3=(\hbar^2/m)/(\rho_*a)$ ($\propto \sqrt{r/a}$), and the equation of state (EoS) in this approximation is given by
\begin{equation}
\frac{E}{N}=\frac{\hbar^2 \rho}{ma}\left[-1+\frac{1}{3}\frac{\rho}{\rho_*}\right].\label{EoS}
\end{equation}
The EoS is shown in Fig.~\ref{fig:EoS}, where it is compared to the results of reference \cite{Bertaina}. The EoS is in qualitative agreement with the Monte Carlo results at the densities where these are available, with the correct order of magnitude. The speed of sound $v$ for $\rho>\rho_*$ can be calculated from the EoS in Eq.~(\ref{EoS}) as $mv^2=\rho\partial_{\rho}\mu$. The Luttinger parameter then has the form
\begin{equation}
K_L= \pi\sqrt{\frac{\rho_*a}{2}}\left(\frac{\rho-\rho_*}{\rho}\right)^{-1/2}.\label{KL}
\end{equation}
Even though the above result -- in particular the factor $\pi/\sqrt{2}$ -- is not accurate quantitatively (see inset of Fig.~\ref{fig:He4}), because higher power terms of $\rho$, and possibly quantum fluctuations \cite{Petrov}, contribute to the chemical potential at such densities, it does show that $K_L$ diverges for $\rho\to \rho_*^+$ as $(\rho-\rho_*)^{-1/2}$, whose exponent agrees with Monte Carlo simulation results \cite{Bertaina}, and unveils the otherwise inaccesible factor of $\rho^{1/2}$ that renders the square root dependence dimensionless. If we assume the functional form of $K_L$ given by Eq.~(\ref{KL}), but leave the factor in front as a renormalisable parameter, i.e. $\pi/\sqrt{2}\leftrightarrow C$, then this can be extracted by fitting the Luttinger parameter for $\rho\in [0.035,0.06]\mathrm{\AA}^{-1}$ to the values reported in \cite{Bertaina}. For our calculated value of $\rho_*=0.0296\AA^{-1}$ we obtain $C=1.388$, while for reference \cite{Bertaina}'s estimate $\rho_*=0.026\AA^{-1}$, we obtain $C=1.697$. In the inset of Fig.~\ref{fig:He4}, we show the resulting renormalised $K_L$, which strongly suggests that the critical point of reference \cite{Bertaina} is most accurate, and that the functional dependence in Eq.~(\ref{KL}) is indeed correct. 

\paragraph{Conclusions}. In this Letter, we have devised a remarkably simple method that uses only two-particle scattering data, that is, the phase shifts, to obtain non-perturbative approximations to the speed of sound and the Luttinger parameter of one-dimensional quantum many-body systems. The method reveals how strikingly large an amount of information the two-body problem in one dimension encodes about the low-energy physics of the many-body system near the Fermi points. We have given relevant examples, such as the Lieb-Liniger model, where third order perturbation theory is recovered, and the speed of sound interpolates deep into the regime where strong-coupling perturbation theory fails. For the simplest case of the hard-rod model our method is, moreover, exact. We have also applied the method to all isotopes of helium and hydrogen, finding excellent agreement with full-blown Monte Carlo simulations available in the literature. We have also shown how well simple two-body theory can predict the critical, or spinodal point in one-dimensional $^4\mathrm{He}$, and extracted the effective three-body force that is largely responsible for the liquid-to-Luttinger liquid transition in this system. On the Bose gas side, we have also obtained the critical exponent with which the Luttinger parameter diverges at the spinodal point ($1/2$), in perfect agreement with Monte Carlo calculations. Our method is not restricted to continuous models, but works as well for lattice models, and can be easily extended to systems with discrete translational invariance, i.e. many-body problems in one-dimensional periodic potentials, for which scattering is also characterised by phase shifts \cite{ValienteKuesterSaenz}. We expect variations of our method to also be able to describe electrons with spin or cold atoms with pseudo spin. Of special interest are systems, such as electrons in a strong magnetic field \cite{Furusaki,Braunecker}, where spin-charge separation is destroyed and the microscopic calculation of the low-energy properties is quite challenging. An extension of our theory to higher-dimensional systems with restricted phase space around the Fermi energy, such as materials with Dirac cones, e.g. for the estimation of the renormalised Fermi velocity in graphene \cite{Graphene,Hofmann,Kotov}, would be highly desirable.  More accurate approximations should also be feasible by extending our results to three- and four-particle problems, which may show effects of non-integrability, but would still be manageable from a theoretical and computational point of view.

\acknowledgements{}
We are indebted to G.~E. Astrakharchik, A.~J. Vidal, L. Vranjes Markic and J. Boronat for sharing the Monte Carlo data of references \cite{AstrakharchikHe} and \cite{AstrakharchikH}, and to G. Bertaina, M. Motta, M. Rossi, E. Vitali and D.~E. Galli for providing their Monte Carlo results of reference \cite{Bertaina}. We thank L. Vranjes Markic for useful discussions about the JDW potential and for sharing her interpolated data. We thank B.~Braunecker, J.~Keeling, N.~T. Zinner and L.~G. Phillips for their valuable feedback. We acknowledge support from EPSRC EP/M024636/1.

\bibliographystyle{unsrt}

\pagebreak
\onecolumngrid
\vspace{\columnsep}
\newpage
\begin{center}
\textbf{\large Supplemental Material: Few-Body Route to One-Dimensional Liquids}
\end{center}
\vspace{2cm}
\twocolumngrid

\setcounter{equation}{0}
\setcounter{figure}{0}
\setcounter{page}{1}
\makeatletter
\renewcommand{\theequation}{S\arabic{equation}}
\renewcommand{\bibnumfmt}[1]{[S#1]}
\renewcommand{\citenumfont}[1]{S#1}
\addtolength{\textfloatsep}{5mm}

\section{Lieb-Liniger and Hard-rod models}
The many-body Hamiltonian of a spinless many-body system is given by
\begin{equation}
H=\sum_{i}\frac{p_i^2}{2m}+\sum_{i<j}V(x_i-x_j).
\end{equation}
For the Lieb-Liniger model, $V(x)=g\delta(x)$, the Lieb-Liniger parameter $\gamma$ is given by 
\begin{equation}
\gamma=\frac{\pi m g}{k_F\hbar^2}.
\end{equation}
The Hard-rod model's potential of diameter $a$ is given by $V(x)=\infty$ if $|x|\le a$ and $V(x)=0$ otherwise.

\section{Universality, coupling constants and perturbation theory}
The simplest way to see how the universality hypothesis is implemented is by means of perturbation theory and "pionless" effective field theory (EFT). The non-interacting lowest-energy state at total zero momentum for two fermions in one dimension on a ring of length $L$, is given by
\begin{equation}
\ket{0}=\frac{1}{\sqrt{2}}\left(\ket{-k_F,k_F}-\ket{k_F,-k_F}\right),
\end{equation}
where $k_F=2\pi / L$. Lowest order in perturbation theory gives the energy
\begin{equation}
E(L)\approx\frac{\hbar^2k_F^2}{m}+\frac{1}{L}\left[V(0)-V(2k_F)\right]\equiv \frac{\hbar^2k_F^2}{m}+\frac{1}{L}V_p(k_F,k_F),\label{PT}
\end{equation}
where we have defined the odd-wave interaction $V_p(k,k')=V(k-k')-V(k+k')$. Since we are considering fermions, we shall use the EFT potential expansion
\begin{equation}
V(q)=\sum_{n=1}^{\infty}g_{2n}q^{2n}.\label{EFTpotential}
\end{equation}
We now place the two-body system in a ring of length $L-\delta$, with $\delta>0$. We expand $1/(L-\delta)$ to lowest order in $\delta$, obtaining
\begin{equation}
\frac{1}{L-\delta}\approx \frac{1}{L}+\frac{\delta}{L^2}.
\end{equation}
Above, we identify $1/L=k_F/2\pi$ and $\frac{\delta}{L^2}=q/2$, where $q$ is the target excitation momentum, since each particle, in the non-interacting case, is effectively placed at a momentum $k_F+q/2$. In this case, however, we can safely assume $q\llt k_F$. The {\it renormalised} coupling constants of the system will have changed when decreasing the size of the system. To see this, we use the Cheon-Shigehara model for fermions \cite{CheonShigeharaS}, corresponding to the lowest-order EFT in Eq.~(\ref{EFTpotential}), i.e. $V(q)=g_2 q^2$. The coupling constant can be renormalised perturbatively to lowest order \cite{ValienteS,ValienteZinnerS,CuiS} (notice that $g_2$ here does not refer to $g_2$ in reference \cite{ValienteS}) as
\begin{equation}
g_2=\frac{2}{g}\left(\frac{\hbar^2}{m}\right)^2,
\end{equation}
where $g$ is the interaction strength of the Lieb-Liniger model, as given by the Bose-Fermi duality \cite{CheonShigeharaS}. If we do not change the potential at all, i.e. if we keep $g$ fixed when changing the size of the system, then the difference in energy $\hbar \omega(q)= E(L-\delta)-E(L)$ is not $\propto q$ for small $q$. We need, in view of the universality hypothesis, to fix the relevant dimensionless quantity of the system, i.e. the Lieb-Liniger parameter $\gamma$, which corresponds to setting
\begin{equation}
g_2=\frac{2\pi}{\gamma}\frac{\hbar^2}{m(k_F+q/2)},
\end{equation} 
when the size of the system is $L-\delta$. To this order, we easily obtain now the correct, well-known result \cite{CazalillaS,ValienteS},
\begin{equation}
\hbar \omega(q)\approx \frac{\hbar^2k_F}{m}\left(1-\frac{4}{\gamma}\right)q.
\end{equation}
To all orders of the EFT potential in Eq.~(\ref{EFTpotential}), we obtain the general condition for the length-dependent (or density-dependent) bare coupling constants $g_{2n}(L)$:
\begin{equation}
ng_{2n}(L-\delta)\cdot (k_F+q/2)=g_{2n}(L)k_F.\label{gn}
\end{equation}
The final result for the lowest-order excitation spectrum, to $\mathcal{O}(q)$, is given by
\begin{align}
\hbar\omega(q)&=\left[\frac{\hbar^2k_F}{m}-\frac{1}{2\pi}\sum_{n=1}^{\infty}g_{2n}\left(2k_F\right)^{2n}\right]q\nonumber\\
&=\left[\frac{\hbar^2k_F}{m}+\frac{1}{2\pi}\left(V(0)-V(2k_F)\right)\right]q,
\end{align}
in perfect agreement with the well-known Born approximation of the speed of sound \cite{GiamarchiS}. The above procedure demonstrates, in a simple, well-known setting, the meaning of keeping the dimensionless constants fixed. It can also be viewed as a renormalisation procedure: Eq.~(\ref{gn}) fits the first order constant correction to the ground state and excited state (with momentum $q$) energy shifts if we identify, as we have done, $k_F=2\pi/L$ or $\rho=2/L$. By enforcing this renormalisation condition, our theory will be correct, in the spirit of traditional renormalisation, if it has any predictive power beyond what it is fitted (renormalised) to reproduce. As we have seen in the Letter, this is indeed the case.

\section{Phase shifts of helium and hydrogen}
\begin{figure}[t]
\begin{center}
\includegraphics[width=0.5\textwidth]{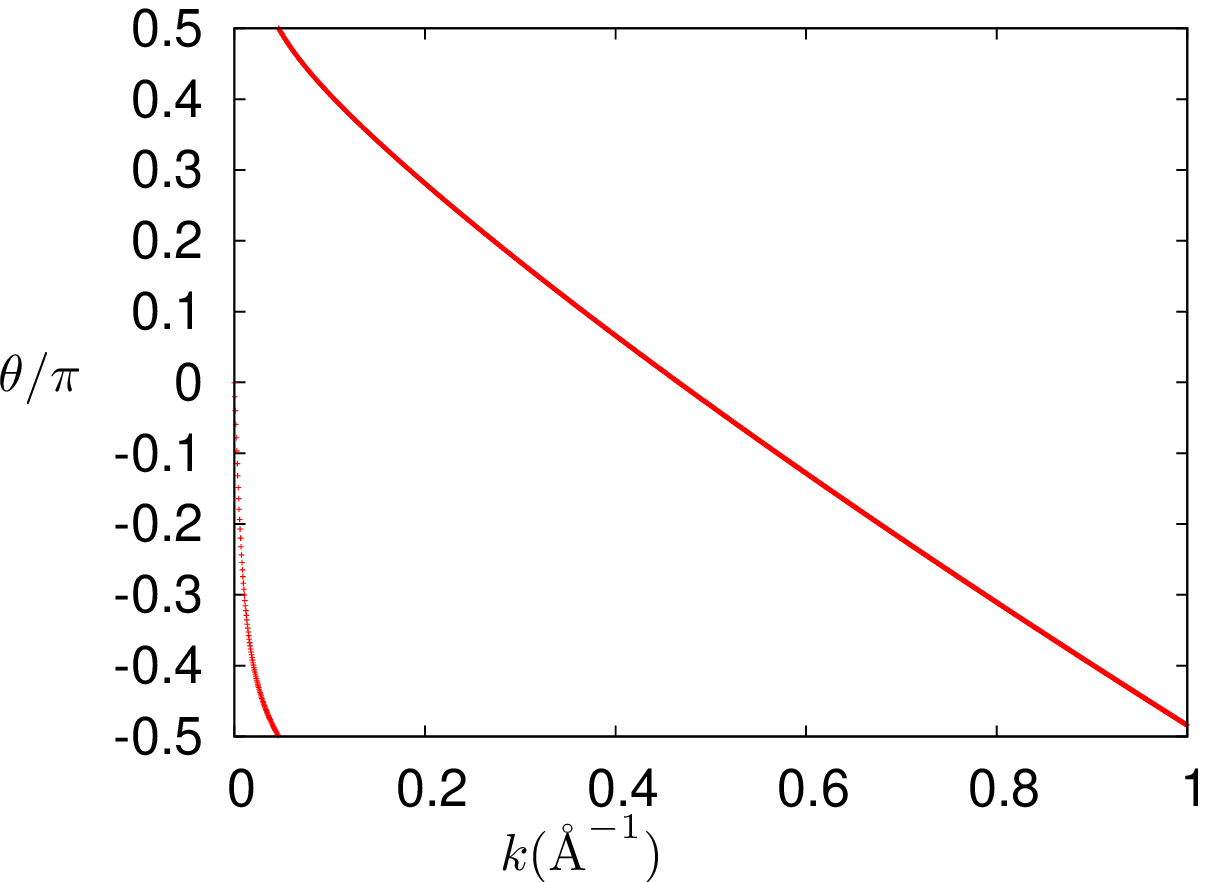}
\end{center}
\caption{Scattering phase shifts for $^4\mathrm{He}$-$^4\mathrm{He}$ collisions using the Aziz HFDHF2 potential.}
\label{fig:phaseshifts-4He}
\end{figure}

\begin{figure}[t]
\begin{center}
\includegraphics[width=0.5\textwidth]{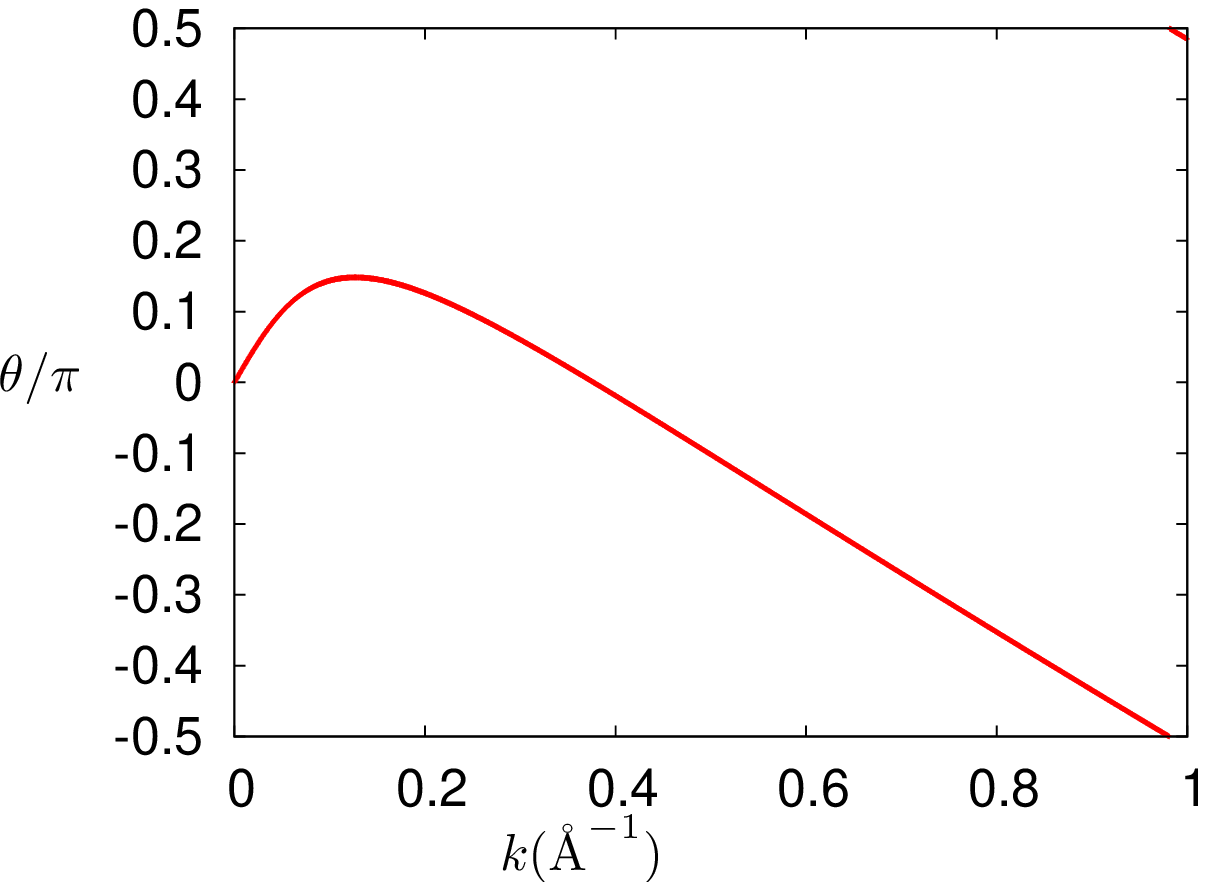}
\end{center}
\caption{Scattering phase shifts for $^3\mathrm{He}$-$^3\mathrm{He}$ collisions using the Aziz II potential.}
\label{fig:phaseshifts-3He}
\end{figure}

As pointed out in the Letter, we use the Aziz HFDHF2 potential for $^4\mathrm{He}$ \cite{AzizHFS}, the Aziz II potential for $^3\mathrm{He}$ \cite{AzizIIS,BoronatAzizS}, and a cubic-spline interpolation of the JDW potential \cite{JDWS}, readjusted to match the $1/|x|^6$ tails \cite{JDW2S} for spin-polarised hydrogen. These are chosen for comparison with available Monte Carlo data, which use these potentials for the respective systems. In order to calculate the phase shifts, and since all these interactions have a short distance hard core, we use the Lippmann-Schwinger equation (LSE) in the position representation for an incident fermionic wave function in the relative coordinate $x=x_1-x_2$. The LSE for the scattering state $\psi_k(x)$ total momentum $K=0$ and energy $E=\hbar^2k^2/m$ reads
\begin{equation}
\psi_k(x)=\sin(kx)+\frac{m}{2\hbar^2k}\int_{-\infty}^{\infty}\mathrm{d}y\sin(k|x-y|)V(y)\psi_k(y).\label{LSE}
\end{equation}
The scattering phase shift $\theta(k)$ is given by
\begin{equation}
\tan\theta(k)=-\frac{m}{2\hbar^2k}\int_{-\infty}^{\infty}\mathrm{d}y\sin(ky)V(y)\psi_k(y).
\end{equation}
We have solved Eq.~(\ref{LSE}) numerically for the different atom-atom interactions and atomic masses, by discretising the integral equation using Gaussian quadrature. Since these potentials have a short-distance hard core and fall off as $\propto 1/|x|^6$ at long distances, we use two sets with an equal number of quadrature points in $(0,r_1)$ and $[r_1,r_2]$, where $r_1$ and $r_2$ are free parameters. We have chosen $r_1=4\mathrm{\AA}$ and $r_2=60\mathrm{\AA}$, which give well-converged results for 2000 quadrature points. The numerically calculated phase shifts for $^4\mathrm{He}$ and $^3\mathrm{He}$ are shown in Figs.~\ref{fig:phaseshifts-4He} and \ref{fig:phaseshifts-3He}, respectively. As is seen in Fig.~\ref{fig:phaseshifts-4He}, the effective interaction is attractive (the scattering length is positive) and near-resonant at low energies, and there is a finite energy resonance at $k=0.0465\mathrm{\AA}^{-1}$.  For $^3\mathrm{He}$, Fig.~\ref{fig:phaseshifts-3He}, and for all the hydrogen isotopes (not shown), the effective interaction is repulsive at low energies. The method described in the Letter is not valid beyond the resonance at $k\sim 1\mathrm{\AA}^{-1}$, where the system is too dense and the short distance hard core will tend to jam the system. 

The locally linear approximation to the phase shifts, $\theta(k)\approx \theta_0-a(k-k_F)$, around a given $k_F$, is obtained by using a forward finite difference with a $5\cdot 10^{-4} \mathrm{\AA}^{-1}$ step size in the momentum.

\section{Speed of sound for light atomic gases}
We obtain here the expression for the speed of sound within the locally linear phase shift approximation, also valid for the hard-rod model. For a size $L-\delta$, we obtain
\begin{equation}
k_{\delta}=\frac{2\pi}{L-\delta}-\frac{2}{L-\delta}\left[\theta_0-a_{\delta}(k_{\delta}-k_F^{(\delta)})\right],
\end{equation}
where we have added a $\delta$-dependence to $a$ and $k_F$ in the phase shift so as to keep $\rho a = 2a/L = \rho_{\delta}a_{\delta}= 2a_{\delta}/(L-\delta)$ constant. In this way, we obtain
\begin{equation}
k_{\delta}=\frac{1}{1-\rho a} \left(\frac{2\pi}{L-\delta}-\frac{2}{L-\delta}\right) \left(\theta_0 + k_F a\right).
\end{equation}
The speed of sound, from Eq.~(\ref{speedofsound}) then gets the form
\begin{equation}
\frac{\hbar^2L^2}{4\pi m}\left(2k_{\delta}\frac{\mathrm{d}k_{\delta}}{\mathrm{d}\delta}\right)\left.\right|_{\delta=0}=\frac{\hbar^2k^2}{mk_F},
\end{equation}
where $k\equiv k_0$. By setting $\theta_0=0$ above, we recover the exact result for the hard-rod model, Eq.~(\ref{vHR}).

 \bibliographystyle{unsrt}

\end{document}